\documentclass[letter,
               keeplastbox,   
               ]{jacow}
\usepackage{pdfpages,multirow,ragged2e} %
\makeatletter%
	\ifboolexpr{bool{xetex}}
	 {\renewcommand{\Gin@extensions}{.pdf,%
	                    .png,.jpg,.bmp,.pict,.tif,.psd,.mac,.sga,.tga,.gif,%
	                    .eps,.ps,%
	                    }}{}
\makeatother

\ifboolexpr{bool{xetex} or bool{luatex}} 
 {}                                      
 {\usepackage[utf8]{inputenc}}           

\usepackage[USenglish]{babel}

\ifboolexpr{bool{jacowbiblatex}}%
 {%
  \addbibresource{jacow-test.bib}
  \addbibresource{biblatex-examples.bib}
 }{}
\usepackage{physics, hyperref}

\begin{document}

\title{low-alpha operation of the iota storage ring}

\author{M. Wallbank\thanks{wallbank@fnal.gov}, J. Jarvis, Fermi National Accelerator Laboratory, Batavia, IL, USA}
	
\maketitle

\begin{abstract}
  Operation with ultra-low momentum-compaction factor (alpha) is a desirable capability for many storage rings and synchrotron radiation sources.  For example, low-alpha lattices are commonly used to produce picosecond bunches for the generation of coherent THz radiation and are the basis of a number of conceptual designs for EUV generation via steady-state microbunching (SSMB).  Achieving ultra-low alpha requires not only a high-level of stability in the linear optics but also flexible control of higher-order compaction terms.  Operation with lower momentum-compaction lattices has recently been investigated at the IOTA storage ring at Fermilab.  A procedure for lowering the ring compaction using the linear optics along with compensations from the higher-order magnets was developed with the aid of a model, and an experimental technique for measuring the momentum compaction was developed.  The lowest momentum compaction achieved during the available run-time was $3.4\times10^{-4}$, around 15 times lower than previously operated.  These feasibility studies ensure an improved experimental understanding of the IOTA optics and potentially will enable new research programs at the facility.
\end{abstract}

\section{INTRODUCTION}

The Integrable Optics Test Accelerator (IOTA)~\cite{IOTA} is a 40\,m storage ring at the Fermilab Accelerator Science and Technology (FAST)~\cite{FAST} facility, designed for flexible beam R\&D.  The research program includes exploring Non-linear Integrable Optics, beam cooling, space-charge effects, single-electron studies and AI/ML for machine operations.  The ring may be injected with electrons accelerated to energies of $50-100$\,MeV by a superconducting linac, or with $2.5$\,MeV protons from a duoplasmatron source currently undergoing commissioning.  There are six sextupole families located at different points around the ring, and nine consecutive octupoles in one straight section.

The momentum compaction $\alpha_c$ of a ring describes the variation in orbit length $\Delta C$ relative to the design orbit $C_0$ experienced by particles at momentum deviations $\delta = (p - p_0) / p_0$, where $p$ and $p_0$ are respectively the particle momentum and the reference momentum, which to third order is written:
\begin{equation}
  \Delta C/C_0 = \alpha_1\delta + \alpha_2\delta^2 + \alpha_3\delta^3.
\end{equation}
The leading-order term $\alpha_1$ is determined by the linear optics, and the higher-order terms can be controlled and corrected by successively higher-order magnets. 

Establishing low-alpha operation of the IOTA ring will enable the development of new accelerator lattice designs which could potentially facilitate new areas of research, and additionally improve confidence in the stability of all experimental lattices through a better understanding of the ring optics.  To support these goals, a short experimental feasibility study aiming to reach a compaction of $1.0\times10^{-4}$ and demonstrate control over the three leading-order terms was carried out at the end of IOTA Run4 (150\,MeV electrons) in Fall 2023.

\section{LOW-ALPHA STORAGE RING MOTIVATIONS}

The primary motivations for exploring low-alpha configurations of the IOTA optics were: to support the ongoing Optical Stochasic Cooling (OSC) experimental program~\cite{OSC-CDR}; to investigate improved phase space acceptance for injection or storage; and to enable new research programs which rely on low-alpha lattices as a fundamental requirement.

The OSC lattice is designed to be low-emittance, to increase the effects of the cooling, which naturally leads to a lower momentum compaction.  In the first phase of the experimental program, there were unresolved problems establishing injection into the lattice until the optics were modified to slightly increase the compaction.  Whilst not investigated at the time, it is possible these issues stemmed from higher-order compaction terms left uncompensated by the non-linear optics.  An improved understanding of controlling these higher-order terms will ensure the best possibility of success in future OSC runs.

As the compaction is reduced, the areas of stable longitudinal phase space are modified and new stable fixed points, so-called alpha buckets, are formed at non-zero $\delta$ and at a phase of $\pi$ with respect to the RF bucket.  For a given lattice, there exists a maximally-stable configuration which particles can populate with significantly longer lifetimes.  Accessing these areas of phase space could be utilized in the injection or storage of particles in the IOTA ring to support greater beam currents and lifetimes \cite{RiesThesis}.

Low-alpha lattices are a fundamental requirement for steady-state microbunching (SSMB), a highly active area of research with great promise as a next-generation light source technology \cite{Ratner2010,Deng2021}.  In SSMB, very short substructure is created within a bunch stored in a ring, which persists and is reinforced through successive turns.  Through the use of a radiator placed at a strategic location, it is conceptually possible to create high power, high frequency radiation comparable to that from a free-electron laser with the high repetition rate of a storage ring.

\section{OPTICAL STOCHASTIC CRYSTALLIZATION}

OSC is a state-of-the-art beam cooling technology first demonstrated experimentally at IOTA in 2021 \cite{Jarvis2022}.  It extends the well-established stochastic cooling technique from microwave to optical bandwidths, enabling significantly increased cooling rates.  The `transit-time' method of OSC~\cite{Zolotorev1994} utilizes undulators for both the `pickup' and `kicker' components, which respectively produce radiation from a bunch and enables subsequent downstream interactions between the particles and their radiation.  The radiation contains information about the longitudinal distribution of the bunch particles to enable corrective energy exchanges in the kicker.  Between the undulators, the beam is directed through a dispersive section to convert momentum discrepancies to a longitudinal spread, whilst the light passes through optics to refocus and optionally amplify the radiation.  The system is phased such that a particle at the reference momentum will arrive in time with its own radiation and thus feel no corrective kicks; particles with an energy discrepancy will gain or lose energy to the radiation fields so that they move closer to the reference energy.  Over time and multiple passes through the system, the bunch is effectively cooled.  The cooling rate is dependent on the bandwidth of the system; that is, how narrow the bunch is sampled to ensure particles are not impacted by others spatially nearby in the kicker.  Additionally, the mechanism relies on sufficient randomization of particles between the kicker and the pickup to ensure collective effects are not enforced.  The fundamental mechanism was demonstrated very successfully without the use of an optical amplifier, and a second phase of the program including an amplifier is currently being designed and scheduled to operate in 2025.

\begin{figure}
  \centering
  \includegraphics[width=0.95\linewidth]{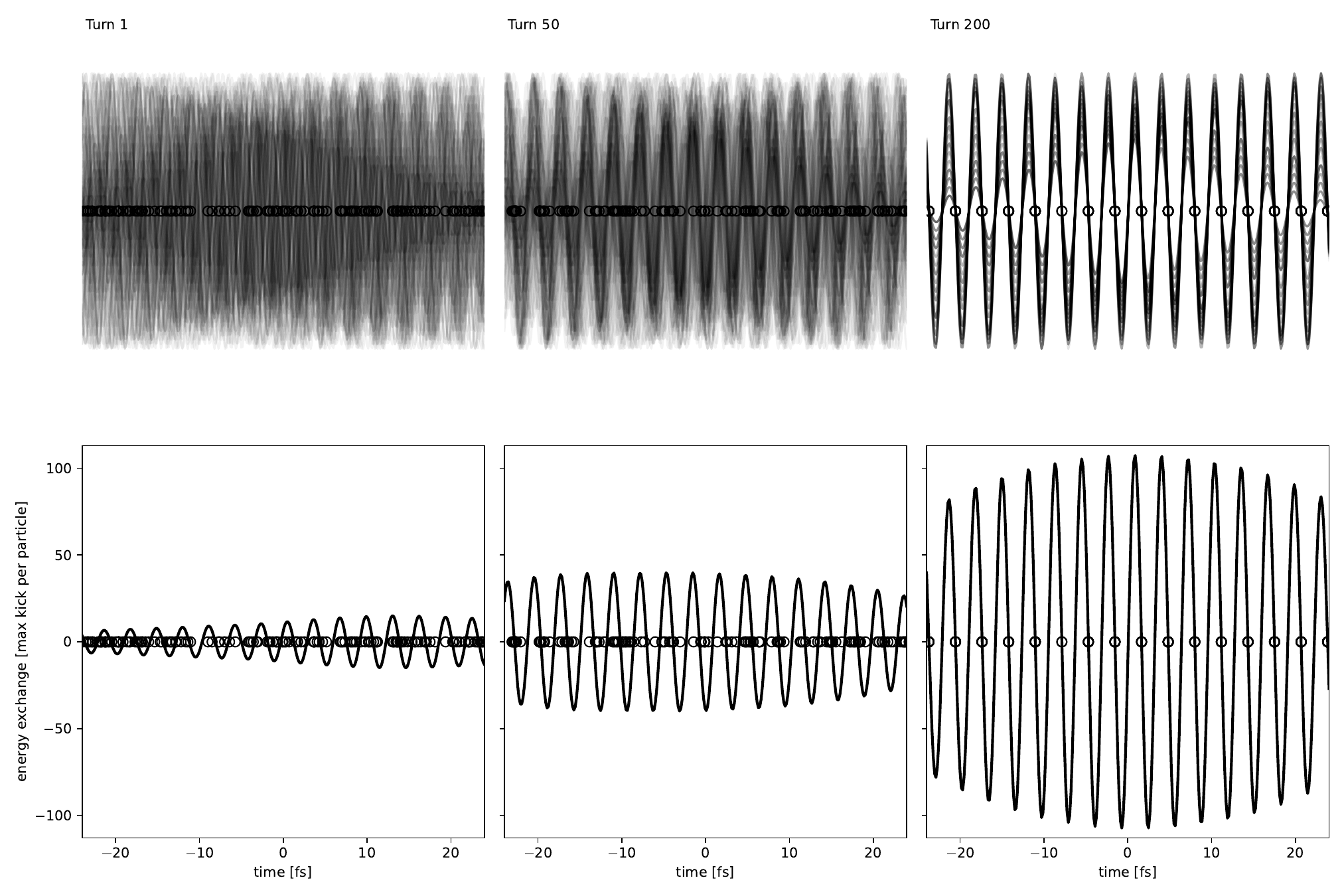}
  \caption{Demonstration of the OSX mechanism using a simple 1D simulation, with three snapshots shown left to right.  The top panels show the radiation produced by initialized randomized particles, and the superposition of all the radiation is shown in the bottom panels.}
  \label{fig:OSXDemonstration}
\end{figure}

An OSC system can be operated to produce longitudinal structure at the optical wavelength, which can lead to SSMB with a mutually configured storage ring; we refer to this as Optical Stochastic Crystallization (OSX).  The mechanism works by instead ensuring insufficient randomization of particles, which leads to strong, self-reinforcing collective effects due to each particle feeling the wake of all neighbors within the system bandwidth.  This is demonstrated in Figure~\ref{fig:OSXDemonstration}.

\begin{figure}
  \centering
  \includegraphics[width=0.95\linewidth]{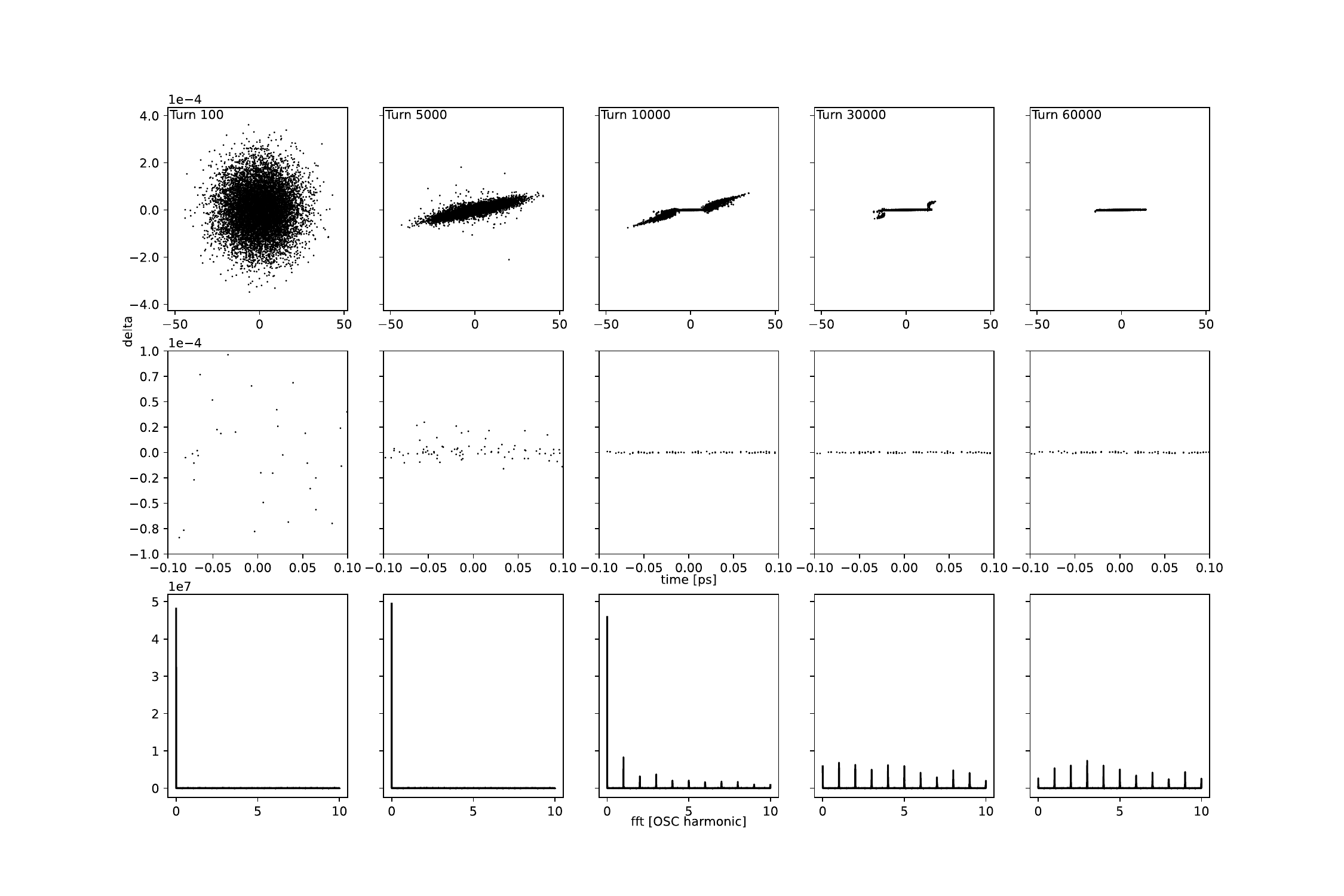}
  \caption{Snapshots from a simple 1D simulation demonstrating the formation of microstructure at the optical wavelength.  The top two panels show the longitudinal phase space, and the power components at harmonics of the optical frequency are shown in the lower panels.}
  \label{fig:OSXSimulation}
\end{figure}

To promote and sustain SSMB via OSX within a storage ring, the following is required: sufficient gain in the OSC system from optical amplification to initiate the formation of substructure; sufficiently low momentum compaction with the same sign as the OSC bypass; and minimal transverse-longitudinal coupling at the undulators, for example by ensuring the dispersion invariant is sufficiently small.  Results from a simple 1D simulation is shown in Figure~\ref{fig:OSXSimulation}, demonstrating the formation of the substructure over successive passes through the ring.  A first experimental demonstration of this concept is planned during the next OSC experimental run in 2025.

\section{LOW-ALPHA DEMONSTRATIONS AT IOTA}

\subsection{Compaction Measurements}

During the initial studies at IOTA, the momentum compaction was estimated from the synchrotron frequency during an RF detuning scan.  For an applied shift in the RF frequency, $\Delta f_{rf} / f_{rf}$, the synchrotron frequency depends on the various orders of momentum compaction as \cite{Hama1993}
\begin{equation}
  \begin{aligned}\label{eq:SynchFreqCompaction}
    f_{s}^{2} &= \frac{h q_{e} V_{rf} f_{0}^{2} \abs{\eta_{1}\cos{\phi_{s}}}}{2 \pi \beta_{0}^{2} E_{0}} \left[1+\frac{s_{1}}{\eta_{1}}\left(\frac{\Delta f_{rf}}{f_{rf}}\right) + \frac{s_{2}}{\eta_{1}^{2}}\left(\frac{\Delta f_{rf}}{f_{rf}}^{2}\right)\right] \\
    s_{1} &= -\frac{2\eta_{2}-\eta_{1}^{2}}{\eta_{1}}+\frac{1}{\gamma_{0}^{2}} \\
    s_{2} &= \frac{3\eta_{3}\eta_{1}-2\eta_{2}^{2}}{\eta_{1}^{2}}-\frac{\eta_{2}}{\eta_{1}\gamma_{0}^{2}}+\frac{3\gamma_{0}^{2}\beta_{0}^{2}+2}{2\gamma_{0}^{4}},
  \end{aligned}
\end{equation}
where $h$, $V_rf$ and $\phi_s$ are the RF harmonic, voltage and synchronous phase respectively, $f_0$ is the revolution frequency in the ring, $q_e$ is the electron charge, and $E_0$ is the beam energy.  The compaction here is given as the slip factor, $\eta$, which very well approximates $\alpha$ in the highly-relativistic case.

The synchrotron frequency was calculated from the sidebands around the 11th harmonic of the RF frequency, measured online by a spectrum analyzer attached to a wall current monitor.  At low compaction, the AC power line frequencies at harmonics of 60\,Hz complicated the measurement, though could relatively easily be filtered out.  This method was found to work extremely well down to the lowest compactions measured.

\subsection{Low-Alpha Investigations}

The accelerator lattice was adapted from the primary version used during IOTA Run4, with an initial momentum compaction of $1\times10^{-2}$.  A strategy to reduce the compaction with the linear optics was developed through the use of a model, and the impact of the non-linear magnets on the higher-order terms was also well characterized.

\begin{figure}
  \centering
  \includegraphics[width=0.95\linewidth]{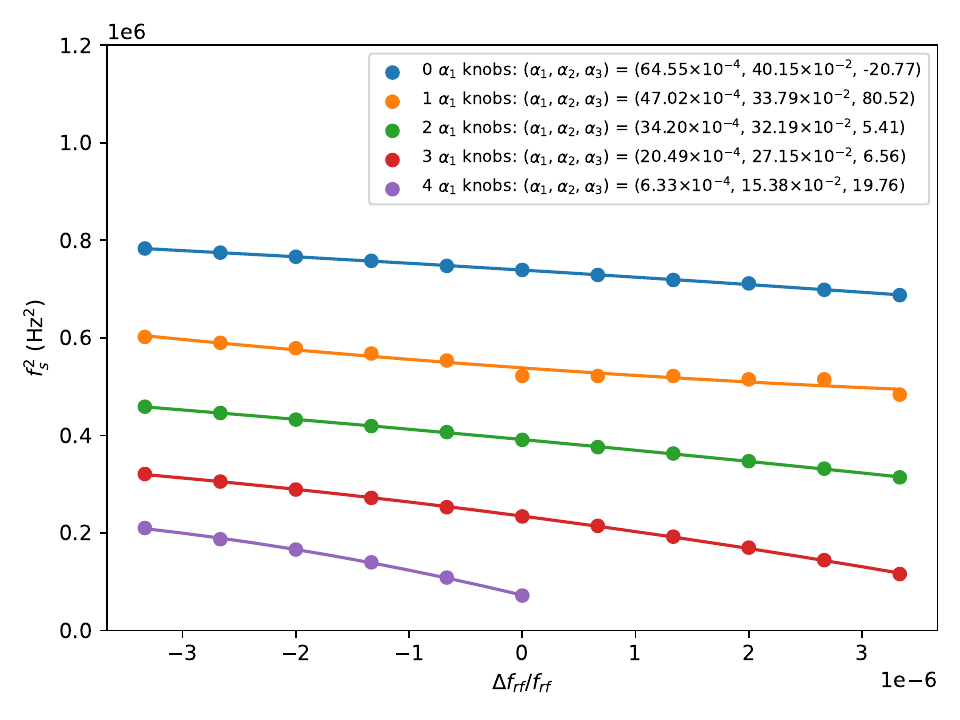}
  \caption{Impact of a linear optics knob designed to reduce the momentum compaction after successive applications.  The leading three $\alpha_c$ terms are shown in the legend.}
  \label{fig:Alpha1}
\end{figure}

The impact of applying successive linear optics knobs, designed to reduce the compaction, is shown in Figure~\ref{fig:Alpha1}.  Each fit represents one scan at a single lattice configuration.  The reduction in the compaction is clear, with the impact of the $\alpha_2$ and $\alpha_3$ terms resulting in a higher-order polynomial relationship becoming visible as $\alpha_c$ is reduced.

\begin{figure}
  \centering
  \includegraphics[width=0.95\linewidth]{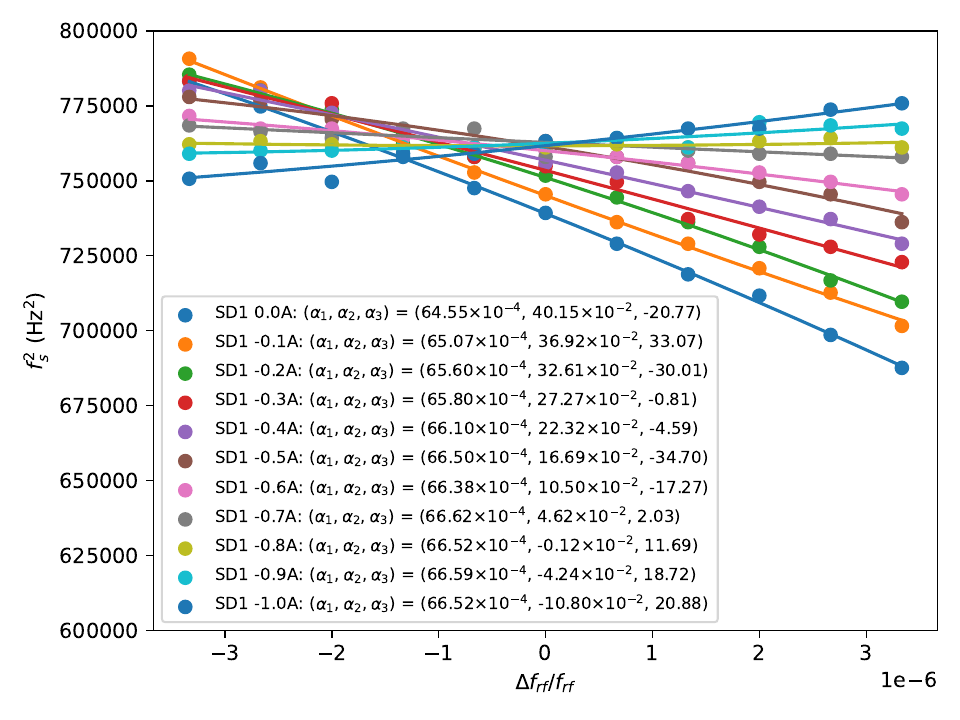}
  \caption{Impact of one sextupoles family (SD1) on $\alpha_c$ at high compaction (no linear optics knobs applied).}
  \label{fig:Alpha2}
\end{figure}

\begin{figure}
  \centering
  \includegraphics[width=0.95\linewidth]{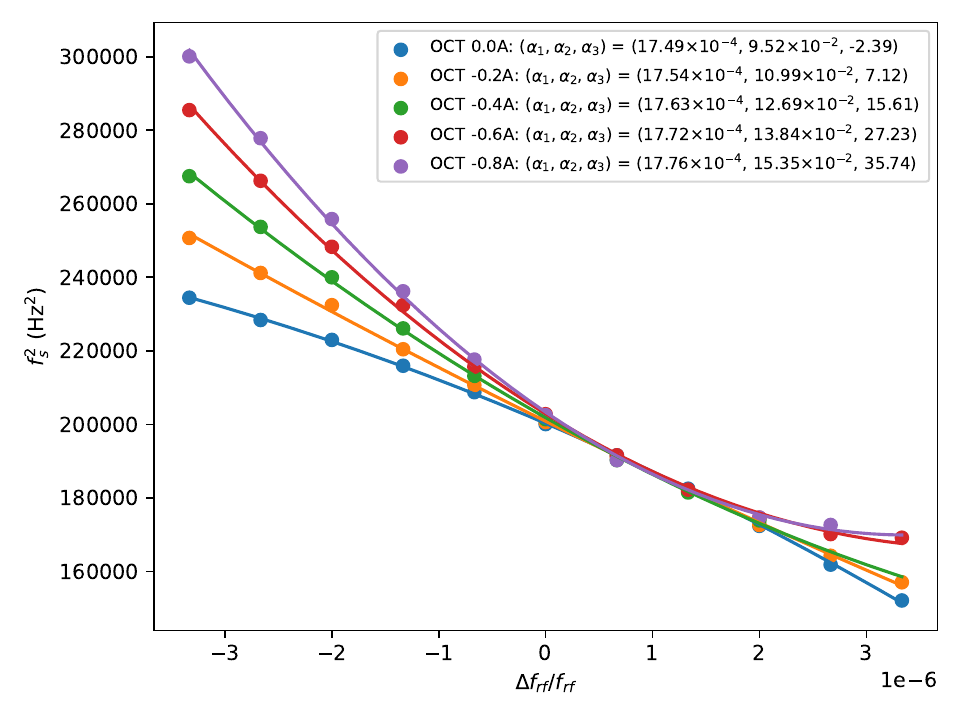}
  \caption{Impact of the octupoles on $\alpha_c$ at lower momentum compaction.  All nine octupoles are energized in series.  For this scan, the sextupoles are used to provide some $\alpha_2$ compensation.}
  \label{fig:Alpha3}
\end{figure}

At high compaction, the ability of the sextupoles to sweep the second-order term $\alpha_2$ through zero is shown in Figure~\ref{fig:Alpha2}.  At lower compaction, resulting from the application of four of the linear optics knobs, the ability of the octupoles to vary the third-order term $\alpha_3$ through zero is shown in Figure~\ref{fig:Alpha3}.  These demonstrations illustrate the required control over the momentum compaction terms from the non-linear optics is acheivable at IOTA.

In order to reach low compactions, the linear optics must be reduced gradually whilst applying the relevant non-linear compensation to ensure stability and sustained lifetime.  Following this procedure, a compaction of $3.4\times10^{-4}$ was reached.  The initial aim of $1.0\times10^{-4}$, and subsequently crossing transition to negative compaction, were not possible given the available experimental time, but the experience gained ensures confidence that these goals are possible.

\section{CONCLUSION}

Low-alpha operation of the IOTA storage has been demonstrated for the first time, validating the requirements on the beam optics to reach ultra-low momentum compactions.  The experience ensures confidence in the next OSC experiments, which includes a first demonstration of Optical Stochastic Crystallization and steady-state microbunching.  The improved understanding of the beam optics will support all ongoing physics programs and potentially allow enhanced standard operational modes and new areas of research.

\section{ACKNOWLEDGMENTS}

We would like to thank Aleksandr Romanov, Nathan Eddy, Daniel Broemmelsiek, and Giulio Stancari for useful discussions and help with detector instrumentation and lattice commissioning.  This manuscript has been authored by Fermi Research Alliance, LLC under Contract No. DE-AC02-07CH11359 with the U.S. Department of Energy, Office of Science, Office of High Energy Physics.

%
%
\ifboolexpr{bool{jacowbiblatex}}%
	{\printbibliography}%
	{%
	
	
} 

%
%


\end{document}